# Deficit of Ultrahigh Energy Extensive Air Showers

A.A. Mikhailov, N.N. Efremov, I.T. Makarov,
G.V. Nikolayeva, G.G. Struchkov

Yu.G. Shafer Institute of Cosmophysical Research
and Aeronomy, 31 Lenin Ave., 677891 Yakutsk, Russia

## Abstract

Arrival directions of extensive air showers by using world data are considered. It is shown that distributions of showers in zenith angle at $E>10^{19}$eV and $E>4\cdot10^{19}$eV differ from each other. On this basis the conclusions are made: the estimation of the energy of inclined showers at $E>4\cdot10^{19}$eV at Yakutsk and AGASA arrays is incorrect; the cosmic ray spectrum is not break up and most likely continued up to $E=10^{20}$eV and higher. It is found that ultrahigh energy cosmic rays are, probably, superheavy nuclei.

## 1. Introduction

Yakutsk [1] and HiRes [2] EAS array data show that the energy spectrum has a break at $E\sim5\cdot10^{19}$eV, and by AGASA data [3] the spectrum is continued without a break up to energies $E=10^{20}$eV and higher. In the solution of this problem the estimation of energy of showers observed is of important part.

## 2. Experiment

Fig.1 presents the distribution of $E>10^{19}$eV showers in zenith angle θ: a – Yakutsk, b – Haverah Park [4]. The number of showers is 458 and 144, respectively. The dashed line is the number of events expected at the observation level [5]. The analysis of the number of showers observed $n_{obs}$ and expected $n_{exp}$ according to the $\chi^2$-criterion shows that at the significance level ~ 0.3 are not contradict each to other in both array data.

Fig.2 shows the distribution of $E>4\cdot10^{19}$eV showers: a – Yakutsk, b – AGASA [6] with the number of showers 29 and 47, respectively. The dashed line is the number of showers expected at the observation level. As seen from Fig.2, a maximum in the distribution of the number of showers is at angle interval of 20-30º and correspondingly the number of inclined showers is less than one expected (see also [7]). The number of showers observed at the Yakutsk array (Fig.2a) according to the $\chi^2$-criterion at the significance level 0.15 is consistent with the number of events expected, but in the case of AGASA at the significance level 0.04 the number of showers observed contradicts to the number of events expected (Fig.2b). If the two shower distribution at Yakutsk and AGASA are joined then according to the $\chi^2$-criterion at the significance level 0.03 the number of showers observed is in contradict with one expected. Thus, the distribution of showers with $E>4\cdot10^{19}$eV in zenith angle is in contradict with the number of events expected at the observation level. Probably, the estimate of the energy of inclined showers at Yakutsk and AGASA arrays requires some correction.

Consider further the distribution of shower in zenith angle based on SUGAR data. Two variants to estimate the energy of showers are given in [8]: by the "Sydney" and "Hillas-E" models. Fig.3 presents the distributions of showers by the "Sydney" model with $E>10^{19}$ eV (a) and with $E>4\cdot10^{19}$ eV (b). The last distribution according to the $\chi^2$-criterion does not agree with the distribution expected. Therefore, the estimation of the shower energy by the "Sydney" model is most likely faulty. By the "Hillas-E" model all the showers in Fig.3a have the energy more than $4\cdot10^{19}$ eV. The distribution of shower at the significance level 0.4 corresponds to the number of events expected. Hence, by the "Hillas-E" model the shower energy is estimated more precisely than by the "Sydney" model. Note that the estimation of the energy of showers with $E>4\cdot10^{19}$ eV by using muon data is most probably more correct than based on the electron-photon component. According to the "Hillas-E" model, at the SUGAR array 8 showers with $E>10^{20}$ eV have been detected. However, in SUGAR data an additional impulse after the passage of the main signal (impulse) in the photomultiplier [8] sets one thinking.

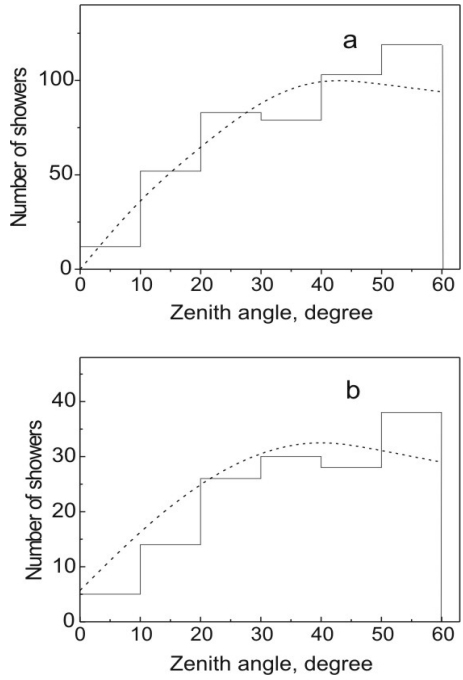
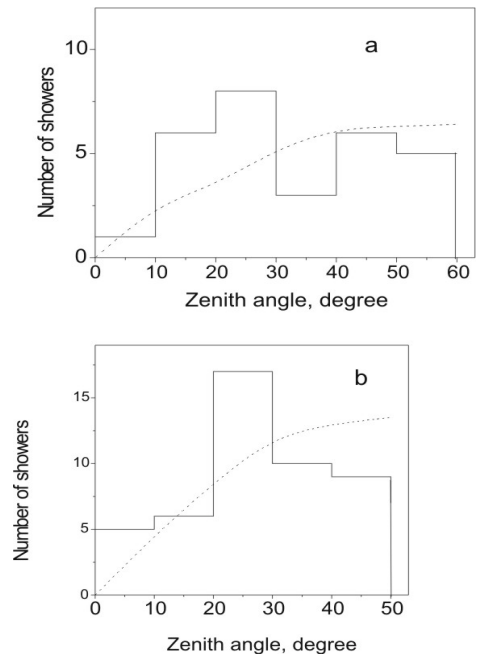

Fig.1. Distribution of showers with $E>10^{19}$ eV over the zenith angle θ according to the (a) Yakutsk and (b) Haverah Park data. The dashed line is the number of showers expected on the level observation.

Fig.2. Distribution of showers with $E>4\cdot10^{19}$ eV over the zenith angle for the (a) Yakutsk and (b) AGASA data. The dashed line is the number of showers expected on the level observation.

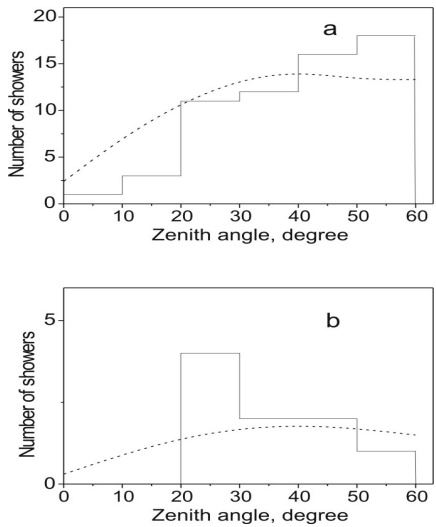

Fig.3. Zenith-angle distribution of showers detected at the SUGAR array (a) with $E>10^{19}$ eV and (b) $E>4.10^{19}$ eV according to the Sydney model and (a) with $E>4\cdot10^{19}$ eV according to the Hillas-E model. The dashed line is the number of showers expected on the level observation.

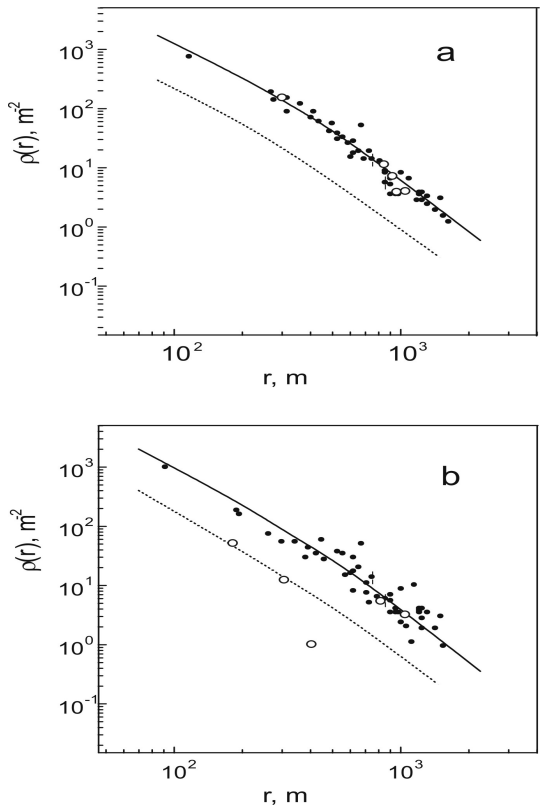

Fig.4. Densities of (closed circles) electron and photon and (open circles) muons at the distance r from the axis of the shower with (a) $E_1=1.2 \cdot 10^{20}$ eV and (b) $E_2=2 \cdot 10^{19}$ eV. The solid and dashed lines are the densities expected for the electron-photon and muons components, respectively.

To clarify reasons of the uncorrected estimation of the shower energy at $E>4 \cdot 10^{19}$ eV we have analyze the muon component of showers by Yakutsk EAS array data. As an example, Fig.4 demonstrates electron-photon and muon components for two inclined showers with $\theta_1=58.7°$, $\theta_2=54.5°$ and $E_1=1.2 \cdot 10^{20}$ eV, $E_2=2 \cdot 10^{19}$ eV. These showers were registered on May 7, 1989 and December 2, 1996 with the axes inside the array perimeter. As seen from Fig.4a, the particle densities in the scintillation detectors (registration threshold for electrons and photons is 3MeV) and muon detectors (registration threshold 1GeV) become equal, i.e. the shower with $E_1=1.2 \cdot 10^{20}$ eV consists of muons only [9]. The shower with the lower energy $E_2=2 \cdot 10^{19}$ eV at such a zenith angle $\theta$ has the electron-photon component (Fig.4b). A fact of increasing a portion of muons in inclined showers with the rise of the energy was established over all the data in [10].

## 3. Discussion

So, at $E>4 \cdot 10^{19}$ eV we have two facts:
1. The number of showers observed is not consistent with one expected at the observation level;
2. In inclined showers the muon component begin to predominate with increasing energy and at $E \sim 10^{20}$ eV it dominates in comparison with the electron-photon component.

The above facts can be interpreted as the change of the cosmic ray mass composition towards more heavy nuclei. The picture of shower development qualitatively is: a heavy nucleus interacts with air atoms relatively high in the atmosphere and the development of shower, possibly is not quite so as at the lower energies (for example, as at $\sim 10^{19}$ eV). Apparently, the electron-photon component of a shower is strongly absorbed (as an example it is the shower with $E=1.2 \cdot 10^{20}$ eV by Yakutsk EAS array data). As the result, the shower energy is estimated by mistake since the shower energy is generally estimated by the particle density at the certain distance from a shower core. Most likely, the shower energy is underestimated, otherwise it is difficult to explain the deficit inclined showers observed at $E>4 \cdot 10^{19}$ eV (Fig.2). The showers underestimated in energy will add to the number of showers with lower energies and probably an excess of showers observed at $\theta>50°$ (Fig.1) is caused by the addition of those showers. The number of muon detectors at the Yakutsk array is insufficient to estimate the shower energy at $E>4 \cdot 10^{19}$ eV by the muon component. The muon detectors at the AGASA array are not capable to measure large densities and the estimation of shower energy at $E>4 \cdot 10^{19}$ eV by them is impossible [11]. So, the underestimation of the shower energy had been traced to a cause of untaking into account in the change of cosmic ray chemical composition at $E>4 \cdot 10^{19}$ eV.

Note, that from 7 showers with E>$10^{20}$eV of AGASA 6 showers are vertical, θ<36° [51]. The inclined showers are only one for AGASA and most likely the spectrum of cosmic rays continue above $10^{20}$eV.

It was shown [12] that cosmic rays at E~$10^{19}$eV were more likely the iron nuclei. From all the above it follows that cosmic rays at E>4·$10^{19}$eV are more heavy than the iron nuclei. It is not improbable that they are transuranic nuclei and are of galactic origin [13].

**4. Conclusion**

1. The estimation of the energy for inclined shower with E>4·$10^{19}$eV by Yakutsk and AGASA data requires the correction.
2. The cosmic ray spectrum is most likely not break up and continued up to $10^{20}$eV and higher.
3. Ultrahigh energy cosmic rays are, apparently, superheavy nuclei with Z>26.

This work is supported by Russian Foundation for Fundamental Research (grant N 00-02-16325). The Yakutsk experiment for detecting EAS is supporting by Russian Ministry for Science (grant N 01-30).

**References**

1. Glushkov A.V., Egorova V.P. et al. Proc. 28th ICRC. Tsukuba. 2003. V.1, P.389.
2. Bergman D.R. and High Resolution Fly's Eye Collaboration. Proc. 28-th ICRC. Tsukuba. 2003. V1. P.397.
3. Takedo M, Sakaki N., Honda K. 28th ICRC. Tsukuba. 2003. V.1. P.381.
4. Linsley J., Reid R.J.O., Watson A.A., Wada M. Catalogue of Highest Energy Cosmic Rays. Tokyo. 1980. V.1. 97 p.
5. Efimov N.N., Pravdin M.I., Mikhailov A.A. Proc. 18-th ICRC. Bangalore. 1983. V.2. P.149.
6. Takeda M., Hayashida N., Honda K. et al. Astro-ph/9902239.
7. Mikhailov A.A. Pisma v ZhETF. 2004. V.79. N.4. P.175.
8. Winn M.M., Ulrichs J., Peak L.S. et al. J. Phys. G: Nucl. Phys. 1986. V.12. P.653.
9. Efimov N.N., Egorov T.A., Glushkov A.V. Proc. Int. Symp. on Astroph. Asp. of the Most Energ. Cosm. Rays. Singapore. 1991. P.20.
10. Glushkov A.V., Makarov I.T., Nikiforova E.S. et al. Astropart. Phys. 1995. V.4. P.15.
11. Nagano M., Watson A.A. Reviews of Modern Physics. 2000. V.72. N.3. P.689.
12. Mikhailov A.A. Pisma v ZhETF. 2000. V.72. P.233.
13. Mikhailov A.A. Pisma v ZhETF. 2003. V.77. P.181.